\documentclass[prl,preprint,superscriptaddress,aps,showpacs]{revtex4-1}
\usepackage[]{graphicx}
\usepackage[]{hyperref}
\usepackage[]{amsmath}
\usepackage{braket}
\usepackage{color}

\newcommand{\plusMinus}{$\{-1$, $+1\}$ }
\newcommand{\zeroMinus}{$\{0$, $-1\}$ }
\newcommand{\plusZero}{$\{+1$, $0\}$ }
\newcommand{\ttwostar}{$T_{2}^{*}$ }

\newcommand{\mechRamsey}{$0.45\pm0.05$}
\newcommand{\waypointRamsey}{$0.36\pm0.09$}
\newcommand{\plusRamsey}{$0.92\pm0.02$}
\newcommand{\minusRamsey}{$0.91\pm0.02$}

\newcommand{\mechRamseyDT}{$830\pm40$}
\newcommand{\waypointRamseyDT}{$140\pm50$}
\newcommand{\plusRamseyDT}{$17\pm3$}
\newcommand{\minusRamseyDT}{$350\pm6$}

\begin{document}

\title{Coherent Control of a Nitrogen-Vacancy Center Spin Ensemble with a Diamond Mechanical Resonator}

\author{E. R. MacQuarrie}
\affiliation{Cornell University, Ithaca, NY 14853}
\author{T. A. Gosavi}
\affiliation{Cornell University, Ithaca, NY 14853}
\author{A. M. Moehle}
\affiliation{Cornell University, Ithaca, NY 14853}
\author{N. R. Jungwirth}
\affiliation{Cornell University, Ithaca, NY 14853}
\author{S. A. Bhave}
\affiliation{Cornell University, Ithaca, NY 14853}
\author{G. D. Fuchs}
\email{gdf9@cornell.edu}
\affiliation{Cornell University, Ithaca, NY 14853}

\begin{abstract}

Coherent control of the nitrogen-vacancy (NV) center in diamond's triplet spin state has traditionally been accomplished with resonant ac magnetic fields under the constraint of the magnetic dipole selection rule, which forbids direct control of the $\Ket{-1}\leftrightarrow\Ket{+1}$ spin transition. We show that high-frequency stress resonant with the spin state splitting can coherently control NV center spins within this subspace. Using a bulk-mode mechanical microresonator fabricated from single-crystal diamond, we apply intense ac stress to the diamond substrate and observe mechanically driven Rabi oscillations between the $\Ket{-1}$ and $\Ket{+1}$ states of an NV center spin ensemble. Additionally, we measure the inhomogeneous spin dephasing time ($T_{2}^{*}$) of the spin ensemble using a mechanical Ramsey sequence and compare it to the dephasing times measured with a magnetic Ramsey sequence for each of the three spin qubit combinations available within the NV center ground state. These results demonstrate coherent spin driving with a mechanical resonator and could enable the creation of a phase-sensitive $\Delta$-system within the NV center ground state. 

\end{abstract}
\pacs{}

\maketitle

Spin-based quantum systems typically rely on resonant magnetic fields to drive coherent transitions between different spin states. Although such magnetic driving has been effective, developing alternative modes of control opens new routes for coupling disparate quantum states to form a hybrid quantum system~\cite{soykal2011}. New techniques for manipulating a spin state also naturally extend to new sensing capabilities and an enhanced understanding of how spin systems interact with their environment. 

The spin triplet ground state of the nitrogen-vacancy (NV) center in diamond represents a coherently addressable paramagnetic defect confined within a largely non-magnetic carbon lattice. This creates an excellent laboratory for studying how spin-based quantum systems interact with their environment~\cite{hanson2008} and for exploring new methods of quantum control~\cite{dobrovitski2013}. Studies have shown that NV center spins can be controlled magnetically~\cite{jelezko2004}, optically~\cite{togan2011,yale2013}, electrically~\cite{dolde2011}, and mechanically~\cite{MacQuarrie2013,teissier2014,ovartchaiyapong2014}. The direct spin-phonon coupling that enables mechanical spin control mediated by lattice strain has prompted the experimental development of single-crystal diamond mechanical resonators~\cite{burek2013,MacQuarrie2013,teissier2014,ovartchaiyapong2014} and motivated theoretical calculations showing that this interaction could enable spin squeezing~\cite{bennett2013} and mechanical resonator cooling~\cite{kepesidis2013}. Nonetheless, coherent Rabi driving of NV center spins with a mechanical resonator has not been previously demonstrated. Furthermore, understanding the dynamics of mechanical driving in spin ensembles could have applications in NV center-based sensing and quantum optomechanics where spin-phonon interactions can be enhanced by using a large number of spins. 

Here we use a mechanical microresonator to apply a large amplitude ac stress to a single crystal diamond. Building on recent spectroscopy experiments~\cite{MacQuarrie2013}, we tune the frequency of this stress wave into resonance with the $\Ket{(m_s=)-1}\leftrightarrow\Ket{+1}$ spin transition to mechanically drive Rabi oscillations of an NV center spin ensemble. Using this capability, we measure the inhomogeneous dephasing time for an ensemble of mechanically controlled NV center spin qubits to be $T_{2}^{*}=$~\mechRamsey~$\mu$s and compare this result to $T_2^*$ for magnetically driven qubits constructed from the same NV center ensemble. We find that the mechanically driven \plusMinus qubit coherence is similar to that of a magnetically driven \plusMinus qubit, and these \plusMinus qubits dephase twice as quickly as magnetically driven \zeroMinus or \plusZero qubits. 

\begin{figure}[ht]
\begin{center}
\begin{tabular}{c}
\includegraphics[width=13cm]{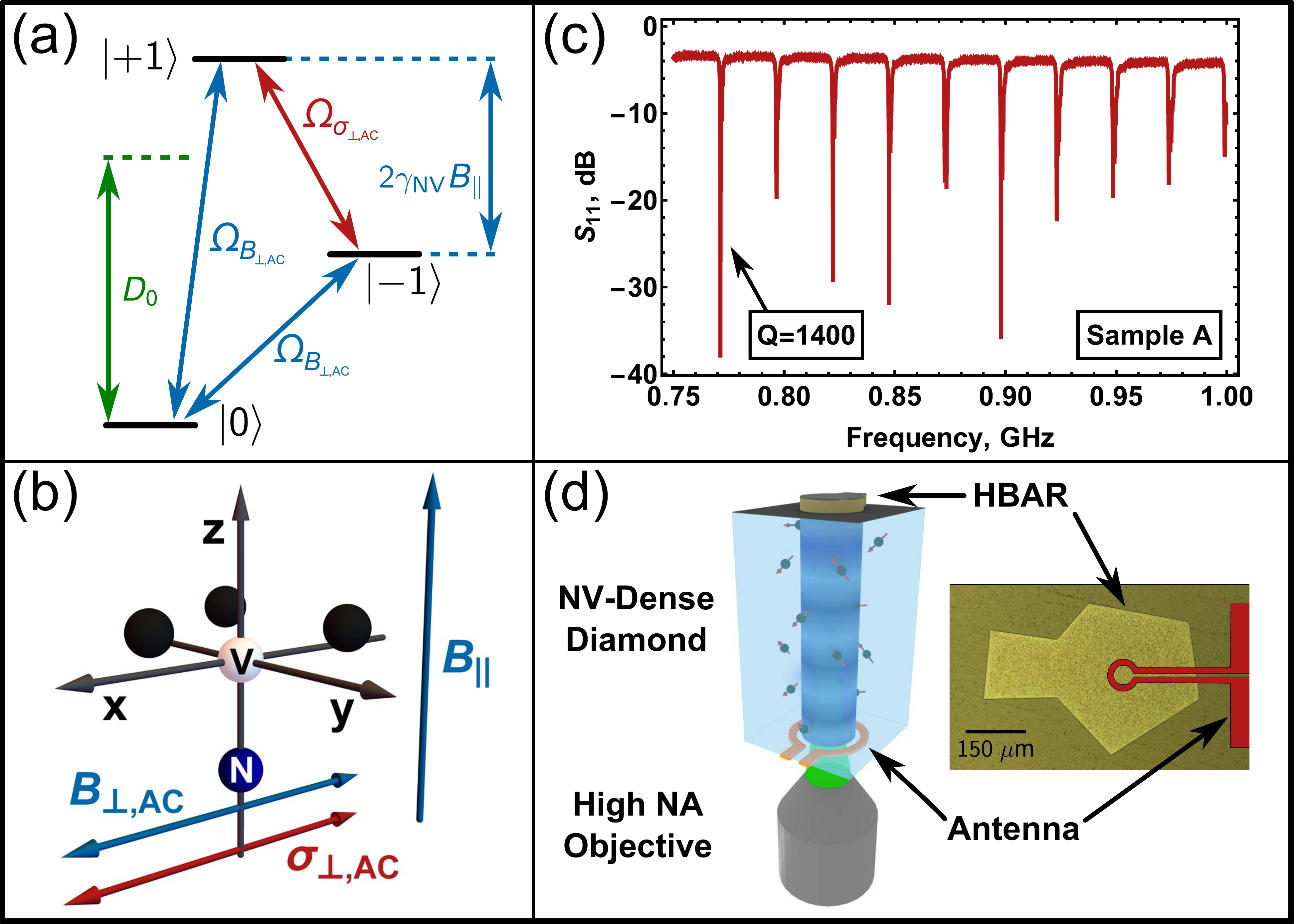} \\
\end{tabular} 
\end{center}
\caption[fig:fig1] {(a) Energy levels of the NV center ground state with corresponding energy separations and driving fields. (b) Schematic of the NV center with applied magnetic ($B_{\bot}$ and $B_{\|}$) and mechanical ($\sigma_{\bot}$) fields. (c) Reflected microwave power ($S_{11}$) as a function of frequency for the Sample~A HBAR as measured with a network analyzer. The resonance at $\omega_{mech}/2\pi=0.771$~GHz has a $Q$ of $1400$. (d) Device schematic (not to scale) and an optical micrograph of an HBAR with the shadow of the loop antenna on the reverse diamond face indicated in red. Apodizing the shape of the HBAR limits the formation of lateral mechanical modes.}
\label{fig:fig1}
\end{figure}

NV centers couple to mechanical stress ($\sigma_{\bot}$ and $\sigma_{\|}$) and magnetic fields ($B_{\bot}$ and $B_{\|}$) through their ground-state spin Hamiltonian (shown schematically in Fig.~\ref{fig:fig1}a) 
\begin{equation}
\begin{split}
H_{NV} &= (D_{0}+\epsilon_{\|}\sigma_{\|})S_{z}^{2} + P I_{z}^{2} + A_{\|} I_z S_z + \gamma_{NV}B_{\|}S_{z} \\
&+ \gamma_{NV} B_{\bot} S_x -\epsilon_{\bot}\sigma_{x}(S_{x}^{2}-S_{y}^{2})+\epsilon_{\bot}\sigma_{y}(S_{x}S_{y}+S_{y}S_{x}) 
\end{split}
\label{eq:Hgs}
\end{equation}
where $D_{0}/2\pi=2.87$~GHz is the zero-field splitting, $\gamma_{NV}/2\pi=2.8$~MHz/G is the gyromagnetic ratio, $\epsilon_{\bot}/2\pi=0.015$~MHz/MPa and $\epsilon_{\|}/2\pi=0.012$~MHz/MPa are the perpendicular and axial stress coupling constants~\cite{ovartchaiyapong2014,SI}, $P/2\pi=-4.945$~MHz and $A_{\|}/2\pi=-2.166$~MHz are the hyperfine parameters~\cite{doherty2012, smeltzer2009, steiner2010}, and {$S_x$, $S_y$, $S_z$} ({$I_x$, $I_y$, $I_z$}) are the $x$, $y$, and $z$ components of the electronic (nuclear) spin-1 operator. The NV center symmetry axis defines the $z$-axis of our coordinate system as depicted in Fig.~\ref{fig:fig1}b. In the Supplementary Information (SI), we use the stiffness matrix for diamond to calculate $\epsilon_{\bot}$ and $\epsilon_{\|}$ from the strain coupling constants $d_{\bot}/2\pi=21.5$~GHz/strain and $d_{\|}/2\pi=13.3$~GHz/strain measured by Ovartchaiyapong, \textit{et al}~\cite{SI, ovartchaiyapong2014}. Non-axial stress $\sigma_{\bot}$ couples the $\Ket{-1}$ and $\Ket{+1}$ spin states, enabling coherent control of the magnetically-forbidden $\Delta m_{s}=\pm 2$ spin transition and providing direct access to the \plusMinus spin qubit. This qubit combination has recently become a topic of interest because it is isolated from thermal fluctuations~\cite{fang2013} and can make a more sensitive magnetometer than either the \zeroMinus or \plusZero qubit~\cite{fang2013, mamin2014}. 

In this work, we use two devices, both fabricated from type IIa, $\langle 100\rangle$ ``optical grade" diamonds purchased from Element Six. These samples are specified to contain fewer than 1 ppm nitrogen impurities, and each contained a native NV ensemble as received. The first sample, Sample~A, has an NV center density of $\sim110$~NVs/$\mu$m$^{3}$, while Sample~B has a density of $\sim120$~NVs/$\mu$m$^{3}$. To generate the large amplitude, high-frequency stress waves needed for coherent mechanical control, we fabricate high-overtone bulk acoustic resonators (HBARs) that use these single crystal diamonds as resonant cavities. The HBARs used for these measurements consist of either a $1.8$~$\mu$m (Sample~A) or a $2.5$~$\mu$m (Sample~B) zinc oxide (ZnO) piezoelectric film sandwiched between a patterned Al electrode and a Ti/Pt ground plane, all sputtered on one face of the diamond substrate. By driving an HBAR with a high-frequency voltage, we transduce stress waves inside the diamond. The diamond then acts as an acoustic Fabry-P\'{e}rot cavity to create standing wave resonances. Fig.~\ref{fig:fig1}c shows a network analyzer measurement of the microwave power reflected ($S_{11}$) from the Sample~A HBAR with the $\omega_{mech}/2\pi=771$~MHz mode ($Q=1400$) used in these experiments indicated. Measurements on Sample~B used a $\omega_{mech}/2\pi=529$~MHz resonance with a $Q$ of $4000$. On the reverse side of each diamond, we fabricate a loop antenna that produces gigahertz-frequency magnetic fields for conventional magnetic spin control. Fig.~\ref{fig:fig1}d depicts a schematic version of the resulting device.

\begin{figure}[ht]
\begin{center}
\begin{tabular}{c}
\includegraphics[width=9cm]{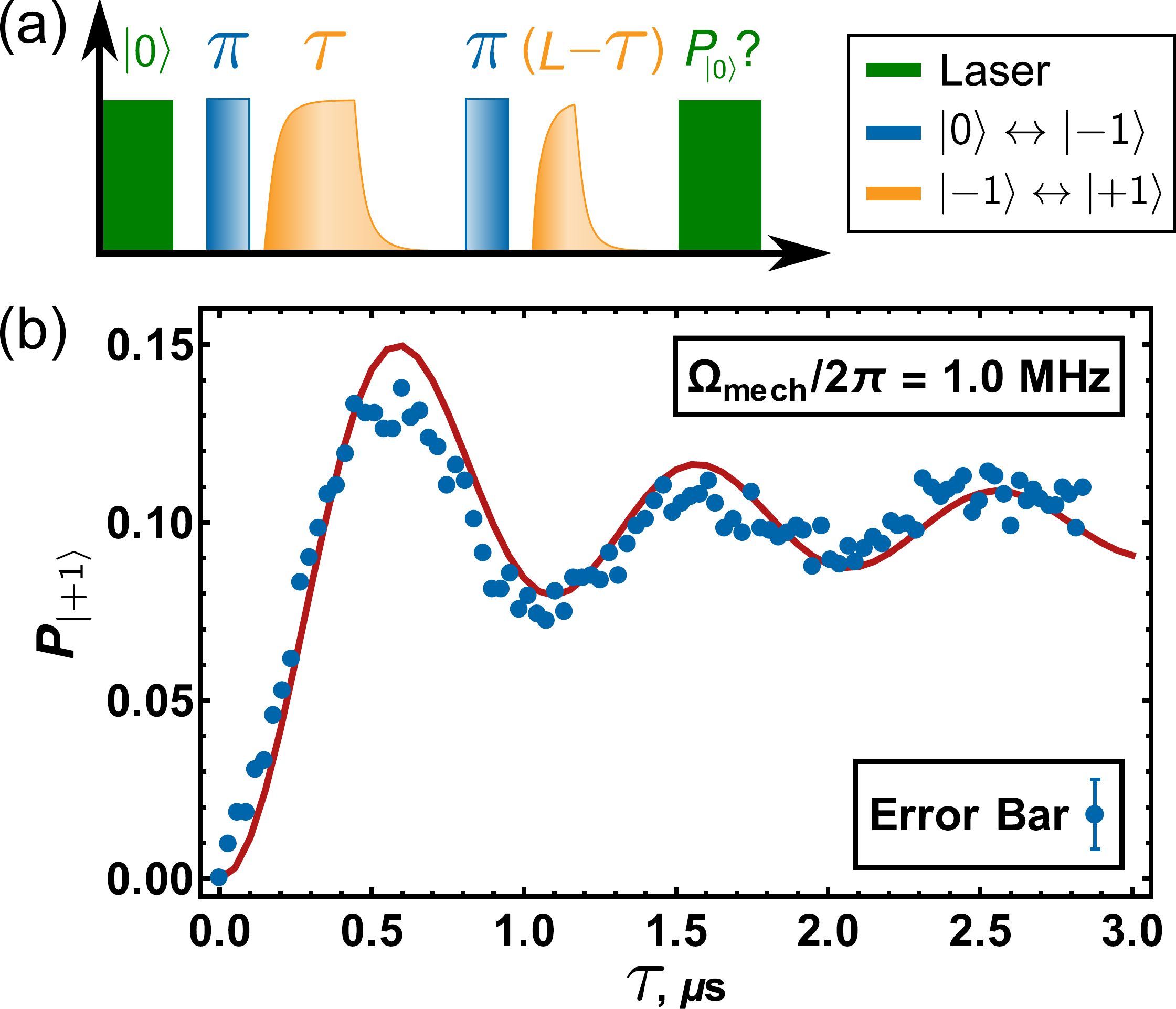} \\
\end{tabular} 
\end{center}
\caption[fig:rabi] {(a) Pulse sequence for mechanical Rabi driving on low $Q$ devices. (b) Mechanically driven Rabi oscillations between the $\Ket{-1}$ and $\Ket{+1}$ spin states for the $\omega_{m}/2\pi=771$~MHz mechanical mode of Sample~A ($Q=1400$). An input power of $33$~dBm produces a Rabi frequency of $\Omega_{mech}/2\pi=1.0$~MHz.}
\label{fig:rabi}
\end{figure}

To perform mechanically driven spin coherence measurements, we first tune the axial magnetic field $B_{\|}$ to bring the spins into resonance with a high-frequency stress wave as described in Ref.~\cite{MacQuarrie2013}. At this resonant $B_{\|}$, we mechanically drive Rabi oscillations of the \plusMinus qubit. Fig.~\ref{fig:rabi}a shows the pulse sequence used to drive Rabi oscillations in the relatively low $Q$ modes of Sample~A. To initialize the NV center spins, we first optically polarize into $\Ket{0}$ and then transfer the spin population from $\Ket{0}$ to $\Ket{-1}$ with a magnetic $\pi$-pulse. Next, we apply a mechanical Rabi pulse of length $\tau$ that is resonant with the $\Ket{-1}\leftrightarrow\Ket{+1}$ spin transition. To read out the spin signal, a second magnetic $\pi$-pulse shuttles the population in $\Ket{-1}$ to $\Ket{0}$. Fluorescence measurement of the $\Ket{0}$ state population reveals how much spin population was transferred to $\Ket{+1}$ according to the relation $P_{\Ket{+1}}=1-P_{\Ket{0}}$~\cite{controlRabiFootnote}. In order to maintain a constant average power to the device, we apply a second mechanical pulse at each data point of length $L-\tau$ where $L$ is the length of the longest Rabi pulse. This pulse comes before fluorescence read out but does not affect our measurement since the spin population we detect has left the \plusMinus subspace. Fig.~\ref{fig:rabi}b shows mechanically driven Rabi oscillations as measured on Sample~A for $33$~dBm of input power to the HBAR. 

The damping observed in Fig.~\ref{fig:rabi}b arises from a combination of spin dephasing from magnetic bath noise and dephasing derived from spatial variations in the amplitude of the stress standing wave within the spin ensemble. NV centers near an anti-node of the stress wave feel a larger Rabi frequency than NV centers near a node. The finite collection volume of our confocal microscope necessitates measuring a distribution of coupling strengths, which causes the measured spin signal to dephase. To account for both of these dephasing sources, we model the data in Fig.~\ref{fig:rabi}b with the spatially-weighted average 
\begin{equation}
\begin{split}
P_{\Ket{+1}}&=\frac{1}{3}\frac{1}{\int_0^\infty \! g(z,z_0) \, \mathrm{d}z} \\
& \times\int_0^\infty \! g(z,z_0)\frac{\Omega(z)^2}{\Omega(z)^2+\delta^2}\sin^2\left[\frac{1}{2} \sqrt{\Omega(z)^2+\delta^2} t\right] \, \mathrm{d}z
\end{split}
\end{equation}
where the factor of $1/3$ arises because we drive only one of the unpolarized nuclear spin sublevels, $\Omega(z)=\Omega_{mech}|\text{sin}\frac{2\pi z}{\lambda_A}|$ is the mechanical driving field, $\lambda_A$ is the wavelength of the stress standing wave, and $g(z,z_0)$ represents a Gaussian approximation to the microscope point spread function (PSF) with a FWHM that grows linearly with the depth of focus inside the diamond $z_0$ as described in Ref.~\cite{MacQuarrie2013}. We assume resonant driving and include quasi-static spin bath noise as a random detuning $\delta$ drawn from a Gaussian distribution with a standard deviation $\sigma=\sqrt{2}/T_2^*$~\cite{aiello2013}. The mechanical Ramsey measurement presented below sets $T_2^*=0.45$~$\mu$s in the \plusMinus subspace. With the parameters $\Omega_{mech}/2\pi=1.0$~MHz, $\lambda_A=19.9$~$\mu$m, and $z_0=18$~$\mu$m as inputs, we average $200$ iterations of the simulation to produce the model curve in Fig.~\ref{fig:rabi}b, which is not a fit to the experimental data.

\begin{figure}[ht]
\begin{center}
\begin{tabular}{c}
\includegraphics[width=13cm]{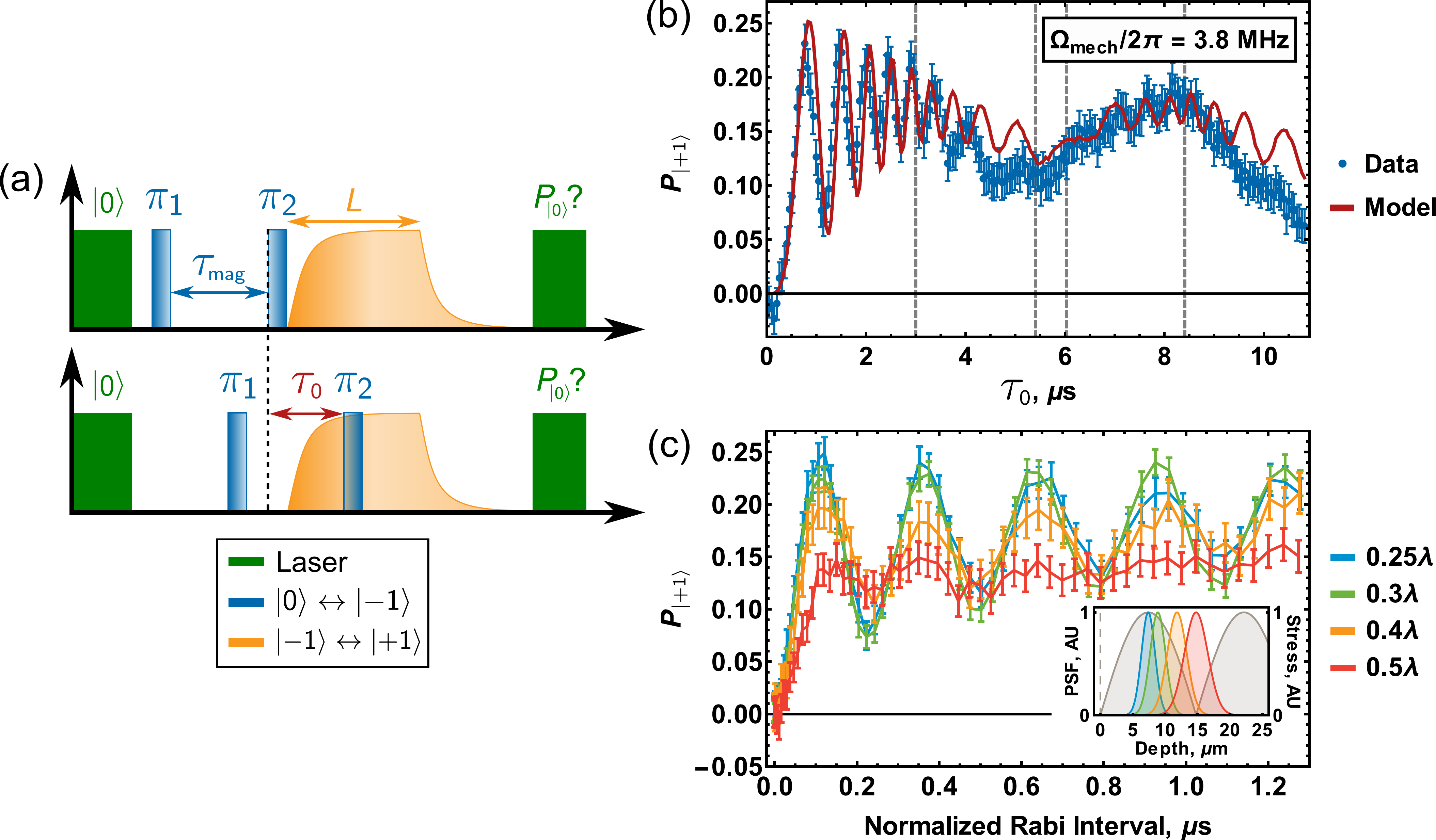} \\
\end{tabular} 
\end{center}
\caption[fig:rabi2] {(a) Pulse sequence for mechanical Rabi driving on high-$Q$ devices. (b) Mechanically driven Rabi oscillations for the $\omega_{m}/2\pi=529$~MHz mechanical mode of Sample~B ($Q=4000$). The model curve is not a fit to the data. From left to right, the dashed lines correspond to $\pi_{2}$ entering the ring down portion of the mechanical pulse, $\pi_1$ entering the ring up, the maximum mechanical pulse area at $\tau_{c}$, and $\pi_1$ entering the ring down. (c) Mechanically driven Rabi oscillations at different depths inside the diamond substrate plotted as a function of the normalized Rabi interval. An input power of $33$~dBm produces a Rabi frequency of $\Omega_{mech}/2\pi=3.8$~MHz.}
\label{fig:rabi2}
\end{figure}

For devices with $Q$-factors substantially larger than Sample~A, we find a standard Rabi pulse sequence is not effective. In these devices, the large bandwidth of short microwave pulses reduces their spectral precision, which in turn distorts the coupling between the mechanical resonator and its microwave drive. This becomes important in the higher $Q$ resonance of Sample~B. To control this effect, we pulse the stress wave for a fixed duration $L$ at each data point. Because the stress wave only affects spins in the \plusMinus subspace, a pair of short ($\sim30$~ns) magnetic $\pi$-pulses separated by a fixed interval $\tau_{mag}$ controls the length of time the mechanical driving field is active. By sweeping this magnetic pulse pair through the mechanical pulse as shown in Fig.~\ref{fig:rabi2}a, we measure mechanically driven Rabi oscillations in the \plusMinus subspace. For $33$~dBm of input power, the mechanical driving field is $\Omega_{mech}/2\pi=3.8$~MHz, which substantially exceeds the dephasing rate~\cite{SI}. 

Fig.~\ref{fig:rabi2}b shows a Rabi measurement using this protocol with the notable transition points in the sweep labeled and described in the figure caption. The model curve in Fig.~\ref{fig:rabi2}b is the average solution of the Schr\"{o}dinger equation for the spin population in $\Ket{+1}$ after being driven by a segment of the mechanical pulse. We model the mechanical pulse with the functions $1-e^{-\frac{t}{\tau_{r}}}$ for ring-up and $e^{-\frac{t-t_0}{\tau_{r}}}$ for ring-down where $t_0=L+\tau_{r}\log(1-e^{-\frac{t}{\tau_{r}}})$ and $\tau_{r}=2Q/\omega_{m}$~\cite{siebert1985}. As before, the model -- which is not a fit to the data -- accounts for driving field inhomogeneities by applying a spatially-weighted average over an approximated optical PSF and includes quasi-static magnetic bath noise through a randomized detuning. The SI provides additional details on the pulse sequence and model~\cite{SI}. 

For the measurement shown, $\tau_{mag}=L+\tau_{r}=5.41$~$\mu$s where $L=3$~$\mu$s. As such, the critical delay $\tau_{c}=6.03$~$\mu$s corresponds to the largest mechanical pulse area enclosed between the two magnetic $\pi$-pulses. To either side of this time step, the pulse area decreases at roughly the same rate. The asymmetry in the data about this point arises because for $\tau_0<\tau_c$ the mechanical pulse amplitude and thus instantaneous driving field is higher than when $\tau_0>\tau_c$. This larger instantaneous driving field offers the spins better protection from magnetic bath noise as evinced by the larger amplitude Rabi oscillations. Our model correctly reproduces this asymmetry, demonstrating the possibility of using a mechanical driving field to achieve continuous dynamical decoupling of an NV center spin from a spin bath~\cite{du2012}.

By modeling the resonator ringing as described above, we can convert the mechanical pulse area between the two magnetic pulses into the ``square-pulse'' units typically used in magnetic Rabi measurements. Fig.~\ref{fig:rabi2}c shows mechanical Rabi oscillations plotted as a function of this normalized Rabi interval for measurements taken at several depths inside the diamond substrate. As expected, the oscillations dephase faster near a node in the stress wave due to driving field inhomogeneities within the ensemble. Near the antinode, however, the relative uniformity of the stress wave mitigates this depth-dependence and, thus, the dephasing from driving field inhomogeneities. 

\begin{figure}[ht]
\begin{center}
\begin{tabular}{c}
\includegraphics[width=14cm]{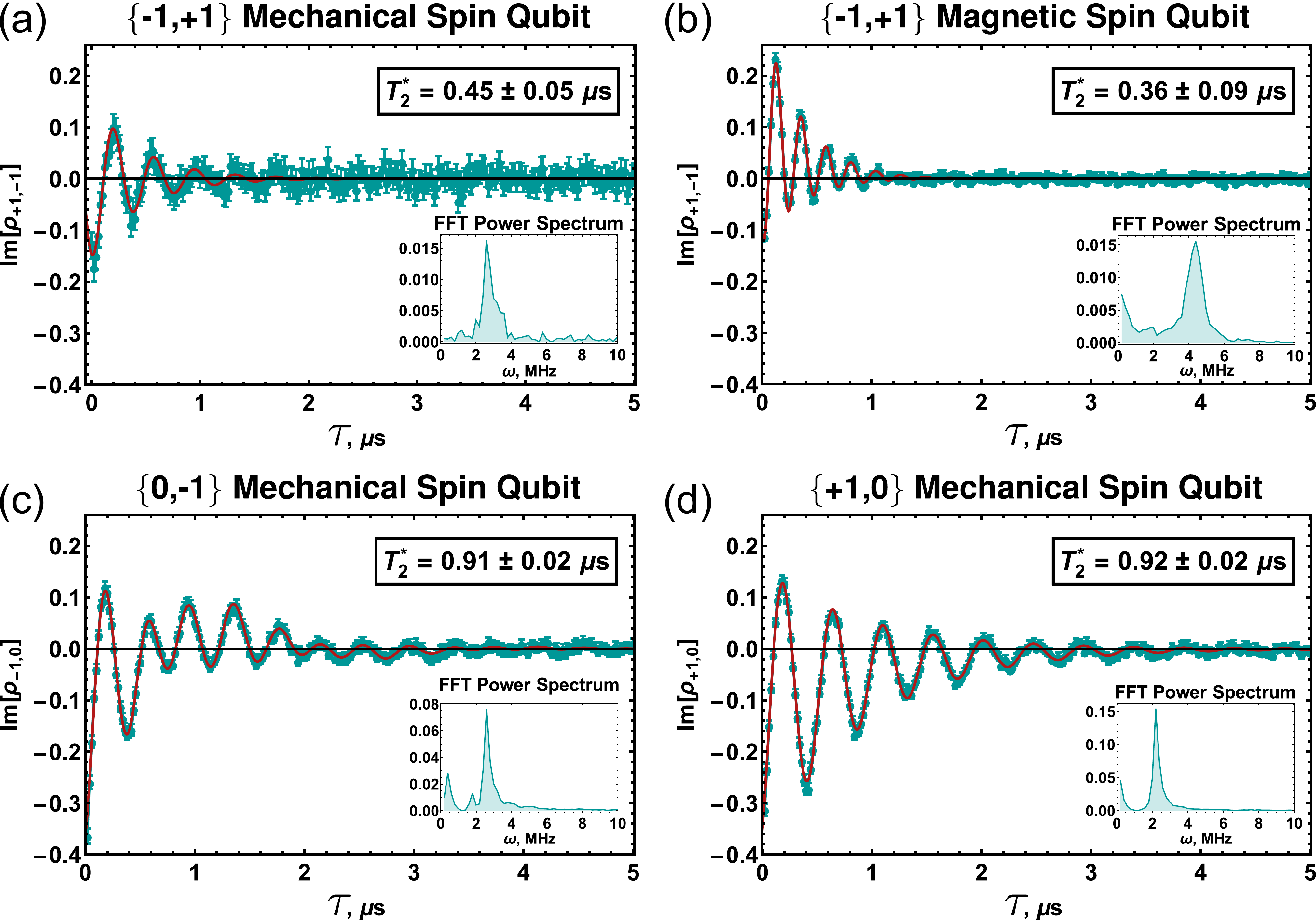} \\
\end{tabular} 
\end{center}
\caption[fig:ramsey] {Ramsey data taken on Sample~A for (a) a mechanically driven \plusMinus qubit ($\delta/2\pi=$\mechRamseyDT~kHz), (b) a magnetically driven \plusMinus qubit ($\delta/2\pi=$\waypointRamseyDT~kHz), (c) a magnetically driven \zeroMinus qubit ($\delta/2\pi=$\minusRamseyDT~kHz), and (d) a magnetically driven \plusZero qubit ($\delta/2\pi=$\plusRamseyDT~kHz).}
\label{fig:ramsey}
\end{figure}

The more traditional Rabi pulse protocol used for Sample~A provides a direct means to implement conventional pulse sequences. From the data in Fig.~\ref{fig:rabi}b, we extract the $\pi/2$-pulse time and proceed to measure $T_{2}^{*}$ of Sample~A with a mechanical Ramsey pulse sequence. Fig.~\ref{fig:ramsey} shows the result of this measurement along with Ramsey measurements of $T_2^*$ for magnetically driven $\{-1$, $+1\}$; $\{0$, $-1\}$; and $\{+1$, $0\}$ qubits. Details on the pulse sequences used for each of these measurements are provided in the SI~\cite{SI}. Although selection rules forbid direct magnetic control of the $\Ket{-1}\leftrightarrow\Ket{+1}$ transition, magnetic control of the \plusMinus qubit can be accomplished indirectly by using either double-quantum pulses~\cite{mamin2014} or multi-pulse sequences~\cite{huang2011}. Both of these alternatives use the $\Ket{0}$ state as a waypoint in the $\Ket{-1}\leftrightarrow\Ket{+1}$ transition. To control the \plusMinus qubit magnetically, we employ the multipulse sequence described in the SI~\cite{SI}.

We fit the three magnetically driven Ramsey measurements to the function
\begin{equation}
\begin{split}
\text{Im}[\rho_{ij}]&=e^{-t/T_{2}^{*}}\{C_1 \cos[(\delta+A_{\|})t+\phi_1]
\\
&+C_2 \cos[\delta t +\phi_2]+C_3 \cos[(\delta-A_{\|})t+\phi_3]\}
\end{split}
\label{eq:ramsey3}
\end{equation}
where $\delta$ represents a detuning in the driving field, the amplitudes ($C_1$, $C_2$, $C_3$) allow for partial polarization of the nuclear sublevels, the constant phases ($\phi_1$, $\phi_2$, $\phi_3$) account for pulse phasing errors, and $A_{\|}\rightarrow 2 A_{\|}$ for the magnetically driven \plusMinus qubit. Since the mechanical driving field ($\Omega_{mech}/2\pi=1.0$~MHz) does not overcome the hyperfine spacing ($2A_\parallel = 4.332$~MHz in the \plusMinus subspace), it drives only one of the nitrogen nuclear spin sublevels. Therefore, we fit our mechanical Ramsey data to the function
\begin{equation}
\text{Im}[\rho_{+1,-1}]=e^{-t/T_{2}^{*}}C_1 \cos[(\delta+\omega_{rot})t+\phi_1]
\label{eq:ramsey1}
\end{equation}
where $\omega_{rot}/2\pi=3.5$~MHz describes an experimentally introduced phase that accumulates at $\omega_{rot}t$ to visualize the decay envelope~\cite{SI}. Our fitting procedure varies $\delta$, $T_{2}^{*}$, $C_i$, and $\phi_i$ as free parameters. Since we measure the coherence of a spin ensemble, we extract \ttwostar from an exponentially decaying envelope rather than from the Gaussian decay expected for a single NV center~\cite{dobrovitski2008}. Fig.~\ref{fig:ramsey} displays the values of \ttwostar extracted from these fits, and the figure caption lists the measured detunings $\delta$. 

The inset within each plot depicts a Fourier power spectrum of the corresponding data. For the magnetic qubits, the Fourier spectra show one peak at $\omega=\delta$ corresponding to the $\Ket{(m_{I}=)0}_{I}$ nuclear spin state. The magnetic \zeroMinus ($\{+1,0\}$) qubit also shows a second peak with roughly twice the amplitude at $\omega_{\pm}= A_{\|}\mp\delta$ ($\omega_{\pm}= A_{\|}\pm\delta$) that represents nearly superposed peaks from the $\Ket{+1}_{I}$ and $\Ket{-1}_{I}$ nuclear states. For the magnetic \plusMinus qubit, this $\Ket{\pm1}_{I}$ peak appears at $\omega_{\pm}=2A_{\|}\pm\delta$. The Fourier spectrum of the mechanical \plusMinus qubit shows only one peak at $\omega=\omega_{rot}+\delta$ because the mechanical driving field drives only one nuclear sublevel. 

For the \plusMinus qubit, we find that \ttwostar measured mechanically (\mechRamsey~$\mu$s) agrees well with \ttwostar measured magnetically (\waypointRamsey~$\mu$s) where the uncertainties equal the square root of the variance in the fitting parameter. The \zeroMinus and \plusZero qubits have dephasing times \ttwostar$=$~\minusRamsey~$\mu$s and \ttwostar$=$~\plusRamsey~$\mu$s, respectively---approximately twice as long as that of the \plusMinus qubit. This agrees with previous measurements performed on a single NV center in low magnetic field~\cite{huang2011,fang2013}. This reduced coherence time does not diminish the \plusMinus qubit's metrological utility because this qubit accumulates phase twice as fast as the longer-lived \plusZero and \zeroMinus qubits, thus reducing the integration time necessary to detect an identical signal~\cite{fang2013,mamin2014}. Additionally, pulsed dynamical decoupling sequences could be implemented in improved devices that take advantage of an anomalous decoherence effect unique to the \plusMinus qubit. This effect can make the spin coherence of the \plusMinus qubit longer than the spin coherence of either the \zeroMinus or the \plusZero qubit decoupled under an equivalent protocol~\cite{huang2011}.

A number of engineering improvements can improve the performance of our devices. First, we expect additional refinements in device fabrication to increase the $Q$ of our devices, which could provide at least a factor of $5$ enhancement in the mechanical driving field~\cite{sorokin2013}. Also, working in higher electronic purity diamond will dramatically reduce spin bath induced dephasing, and working with either a single spin or a plane of NV centers would remove dephasing from driving field inhomogeneities. Taken together, these advances can unlock high fidelity quantum control of a mechanically driven qubit. 

Our results demonstrate coherent control of all three ground state spin transitions. By simultaneously driving the $\Ket{0}\leftrightarrow\Ket{-1}$ and $\Ket{+1}\leftrightarrow\Ket{0}$ transitions magnetically and the $\Ket{-1}\leftrightarrow\Ket{+1}$ transition mechanically, a $\Delta$-system in which all three states are coupled by a closed-loop interaction contour can be created within the NV center ground state. Such a system requires at least one parity non-conserving driving field, making $\Delta$-systems an uncommon extension of the more typical $\Lambda$-system, which has been well explored in NV centers~\cite{hemmer2001, santori2006, santori2006single, togan2011, acosta2013, yale2013}. In a $\Lambda$-system, driving field amplitudes and detunings balance to enable phenomena such as coherent population trapping~\cite{santori2006, santori2006single} and electromagnetically induced transparency~\cite{hemmer2001, acosta2013}. In a $\Delta$-system, similar phenomena occur but with an additional sensitivity to the relative phases of the driving fields~\cite{buckle1986,shahriar1990,kosachiov1992}. Implementing an NV center $\Delta$-system could, for instance, create a phase induced transparency where the phase of a magnetic driving field tunes the absorption of the mechanical driving field. Such a system could have value in NV center optomechanics experiments as a phase-controlled switch to rapidly gate spin-phonon interactions. Another application could be measuring the relative phase of a resonating mechanical proof mass in an inertial sensor. 

In summary, we use a high-frequency mechanical resonator to drive coherent Rabi oscillations of an NV center spin ensemble with driving fields up to $\Omega_{mech}/2\pi=3.8$~MHz. This enabled a comparison of the inhomogeneous dephasing time $T_{2}^{*}$ of a mechanically driven \plusMinus qubit with that of magnetically driven $\{-1$, $+1\}$; $\{0$, $-1\}$; and $\{+1$, $0\}$ qubits. We found that, for both mechanical and magnetic driving, the \plusMinus qubit dephases twice as fast as the \zeroMinus and \plusZero qubits. These results establish the possibility of creating a phase-sensitive $\Delta$-system within the NV center ground state, which could have applications in metrology, optomechanics, and quantum control. 

\vspace*{4mm}

\section{Acknowledgments}
Research support was provided by the Office of Naval Research (ONR). ERM received support from the Department of Energy Office of Science Graduate Fellowship Program (DOE SCGF), made possible in part by  the American Recovery and Reinvestment Act of 2009, administered by ORISE-ORAU under contract no. DE-AC05-06OR23100.  Device fabrication was performed in part at the Cornell NanoScale Science and Technology Facility, a member of the National Nanotechnology Infrastructure Network, which is supported by the National Science Foundation (Grant ECCS-0335765), and at the Cornell Center for Materials Research Shared Facilities which are supported through the NSF MRSEC program (DMR-1120296).

\pagebreak

\section{Supplementary Information}

\section{NV Center Stress Coupling}

Ovartchaiyapong, \textit{et al} measured the NV center strain coupling to be $d_{\bot}/2\pi=21.5$~GHz/strain and $d_{\|}/2\pi=13.3$~GHz/strain for perpendicular and axial strain, respectively~\cite{ovartchaiyapong2014}. Since our mechanical resonator generates acoustic waves by applying a pressure to one face of the diamond crystal, we choose to work in units of stress. To convert the measured constants from strain to stress, we first rotate the measured couplings from the coordinate system defined by the NV center to the lattice coordinates. We then use the stiffness matrix for diamond~\cite{irgens}
\begin{equation}
\begin{pmatrix}
\sigma_{xx}\\
\sigma_{yy}\\
\sigma_{zz}\\
\sigma_{yz}\\
\sigma_{zx}\\
\sigma_{xy}
\end{pmatrix}
=
\begin{pmatrix}
C_{11}	&	C_{12}	&	C_{12}	&	0	&	0	&	0	\\
C_{12}	&	C_{11}	&	C_{12}	&	0	&	0	&	0	\\
C_{12}	&	C_{12}	&	C_{11}	&	0	&	0	&	0	\\
0	&	0	&	0	&	C_{44}	&	0	&	0	\\
0	&	0	&	0	&	0	&	C_{44}	&	0	\\
0	&	0	&	0	&	0	&	0	&	C_{44}	
\end{pmatrix}
\begin{pmatrix}
\epsilon_{xx}\\
\epsilon_{yy}\\
\epsilon_{zz}\\
\epsilon_{yz}\\
\epsilon_{zx}\\
\epsilon_{xy}
\end{pmatrix}
\end{equation}
to convert strain/GHz into GPa/GHz (stress/GHz). The elastic constants $C_{ij}$ are given in Table~\ref{tab:diamondStiffness}. Finally, we rotate back into the coordinates of the NV center to find the stress coupling constants $\epsilon_{\bot}/2\pi=0.015$~MHz/MPa and $\epsilon_{\|}/2\pi=0.012$~MHz/MPa used in the main text. 

\begin{table}[t]
\begin{center}
\begin{tabular}{c | c | c}
$C_{11}$	&	$C_{12}$	&	$C_{44}$	\\
\hline
$1076.4$~GPa	&	$125.2$~GPa	&	$577.4$~GPa 
\end{tabular}
\caption[Stiffness Constants for Diamond]{Stiffness constants for diamond\cite{klein1993}. }
\label{tab:diamondStiffness}
\end{center}
\end{table}

\section{Mechanical Rabi Measurements}

\subsection{Readout Through $\Ket{+1}$}

As a control, we performed a second type of Rabi measurement. In this alternative pulse sequence, after optically pumping the NV center into $\Ket{0}$ we once again apply a magnetic $\pi$-pulse to resonantly move the population from $\Ket{0}$ to $\Ket{-1}$. We then pulse the resonant mechanical driving field for a length $\tau$ to drive the $\Ket{-1}\leftrightarrow\Ket{+1}$ transition. Finally, we use a magnetic adiabatic passage to robustly transfer the population that was driven into $\Ket{+1}$ to $\Ket{0}$ where we read out the spin state optically. This differs from the Rabi measurement presented in the main text in that we extract population from $\Ket{+1}$, not $\Ket{-1}$, for optical readout. 

\begin{figure}[ht]
\begin{center}
\begin{tabular}{c}
\includegraphics[width=9cm]{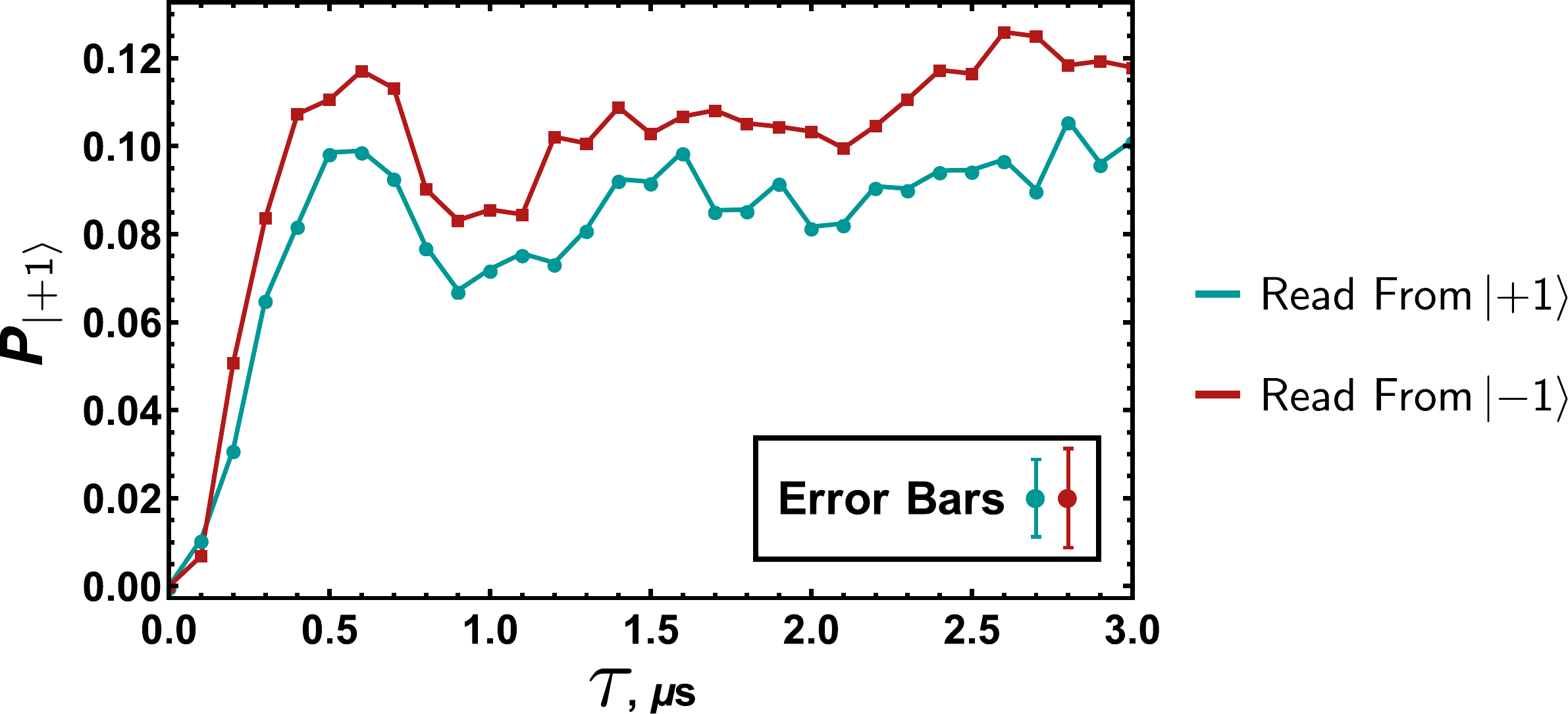} \\
\end{tabular} 
\end{center}
\caption[Control Rabi] {Mechanically driven Rabi oscillations as read out from the $\Ket{+1}$ (blue) and $\Ket{-1}$ (red) spin states. These measurements were performed on Sample~A. }
\label{fig:controlRabi}
\end{figure}

Fig.~\ref{fig:controlRabi} shows the results of this measurement plotted alongside a mechanically driven Rabi measurement that uses a magnetic adiabatic passage to transfer population from $\Ket{-1}$ to $\Ket{0}$ after the mechanical Rabi pulse. Both of these measurements were done on Sample~A. As expected, the results are nearly identical. The difference in amplitudes comes from fidelity differences between the $\Ket{+1}\leftrightarrow\Ket{0}$ and $\Ket{0}\leftrightarrow\Ket{-1}$ magnetic pulses. 

\subsection{Mechanical Rabi Sequence for Sample~B}

\begin{figure}[ht]
\begin{center}
\begin{tabular}{c}
\includegraphics[width=9cm]{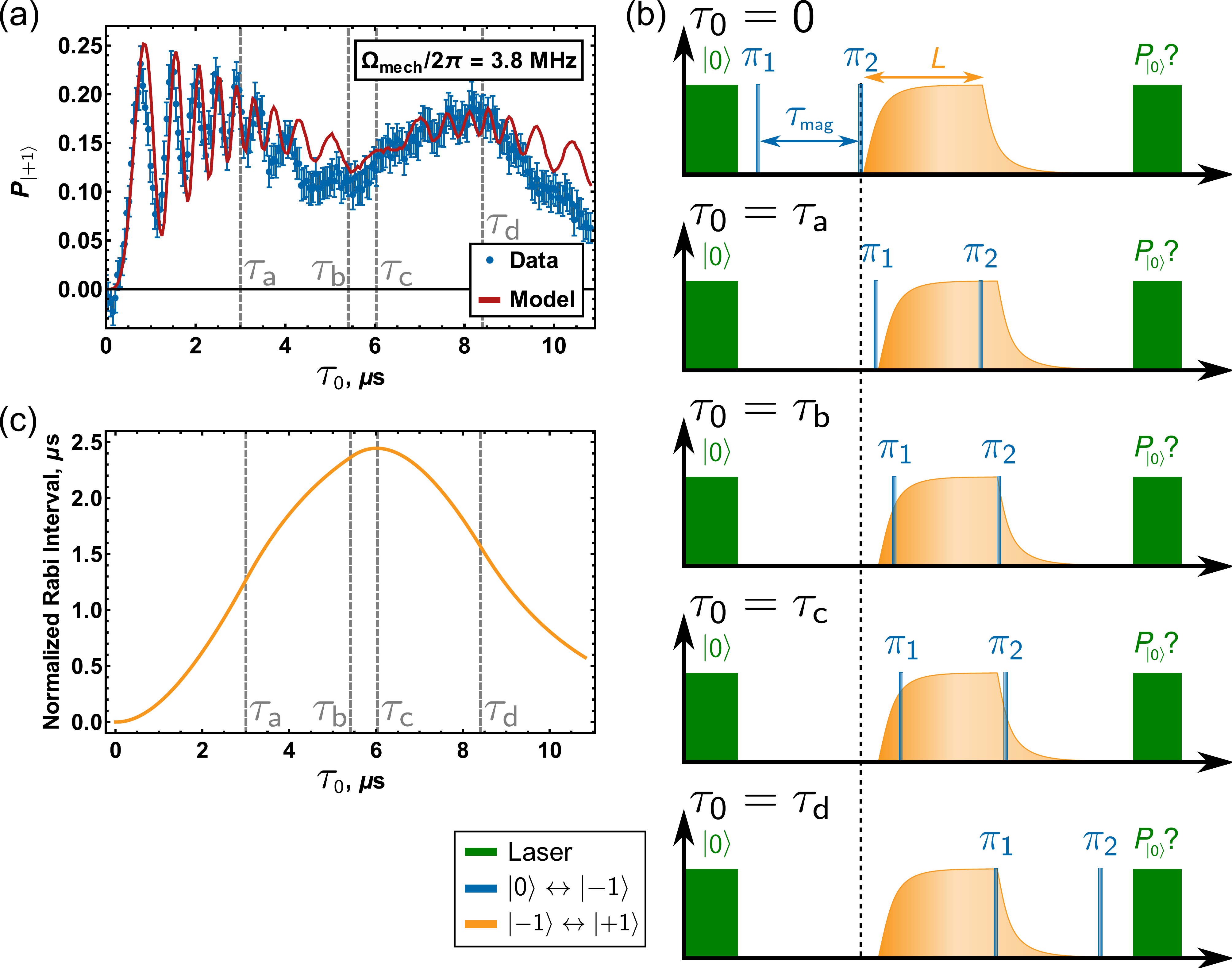} \\
\end{tabular} 
\end{center}
\caption{(a) Rabi oscillations driven mechanically with a high $Q$ mechanical resonator. From left to right, the dashed lines correspond to $\pi_{2}$ entering the ring down portion of the mechanical pulse, $\pi_1$ entering the ring up, the maximum mechanical pulse area $\tau_{c}$, and $\pi_1$ entering the ring down. (b) Pulse sequence at each of the notable times labeled in (a) and (c). (c) Mechanical pulse area enclosed between the two magnetic $\pi$-pulses as a function of $\tau_{0}$. For a mechanical pulse normalized to its amplitude after ring up, this pulse area corresponds to the normalized Rabi interval. }
\label{fig:rabiStepByStep}
\end{figure}

Fig.~\ref{fig:rabiStepByStep}a shows the mechanical Rabi oscillations plotted in Fig.~\ref{fig:rabi2}b of the main text. This measurement was taken by sweeping a pair of magnetic $\pi$-pulses through a fixed-length mechanical pulse. To further elucidate this pulse sequence, Fig.~\ref{fig:rabiStepByStep}b provides a snapshot of the pulse sequence at each of the notable points indicated by dashed lines in Fig.~\ref{fig:rabiStepByStep}a and described in the figure caption. 

We model the ringing of a normalized mechanical driving field with the functions $1-e^{-\frac{t}{\tau_{r}}}$ for ring-up and $e^{-\frac{t-t_0}{\tau_{r}}}$ for ring-down where $t_0=L+\tau_{r}\log(1-e^{-\frac{t}{\tau_{r}}})$ and $\tau_{r}=2Q/\omega_{m}$~\cite{siebert1985}. These functions allow us to compute the mechanical pulse area enclosed between the two magnetic $\pi$-pulses for each value of $\tau_0$. Fig.~\ref{fig:rabiStepByStep}b plots this normalized Rabi interval as a function of $\tau_0$. 

\subsection{Mechanical Rabi Model for Sample~B}

\begin{figure}[ht]
\begin{center}
\begin{tabular}{c}
\includegraphics[width=8.5cm]{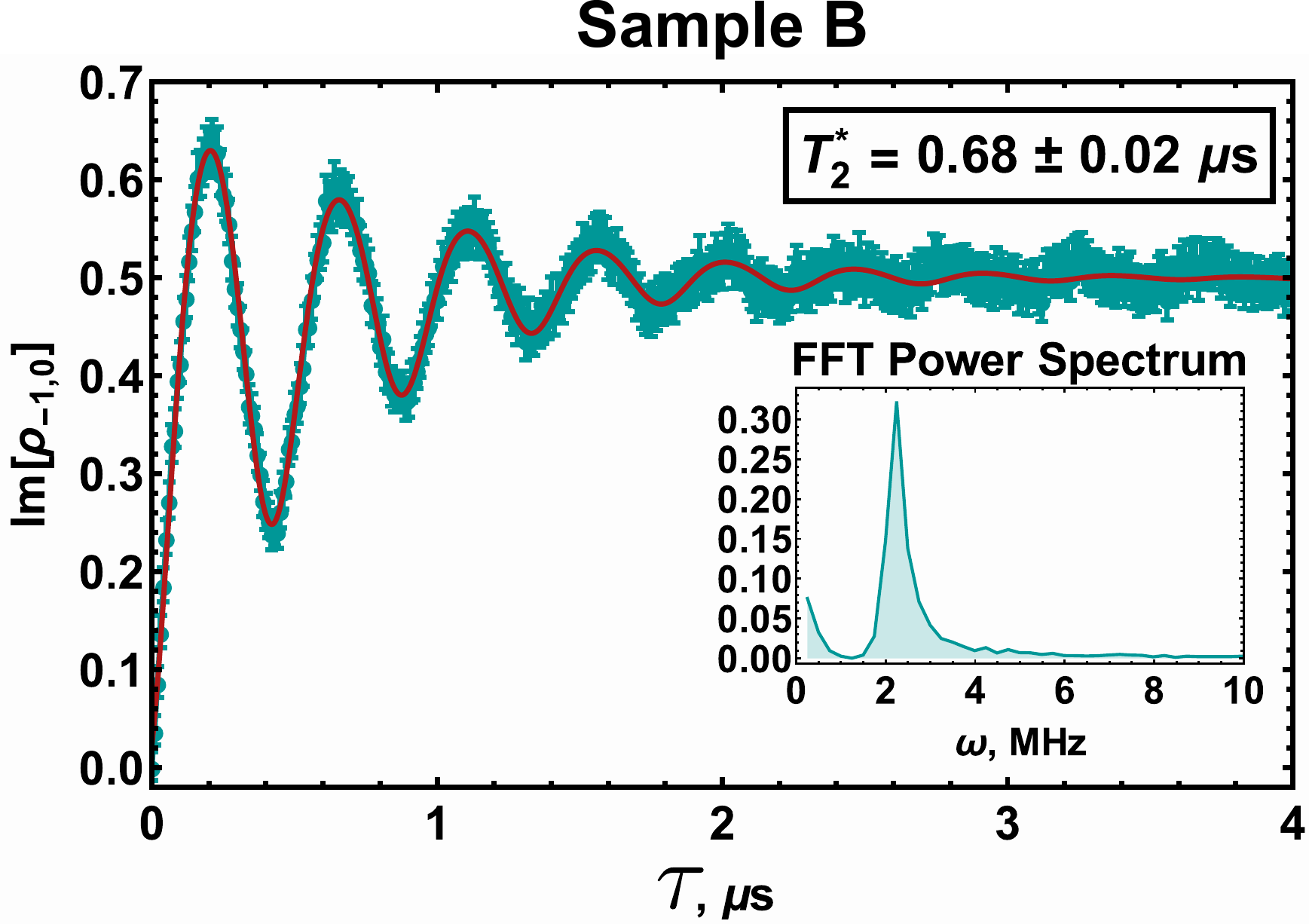} \\
\end{tabular} 
\end{center}
\caption[Ramsey for Sample B] {Magnetic Ramsey measurement of $T_{2}^*$ for Sample~B in the \zeroMinus subspace.}
\label{fig:sampleBRamsey}
\end{figure}

To fit the mechanical Rabi data shown in Fig.~3b of the main text, we solve the Schr\"{o}dinger equation to find the population in $\Ket{+1}$ after applying the relevant portion of an $L=3$~$\mu$s mechanical pulse. We use the Hamiltonian 
\begin{equation}
H_{up}=\begin{pmatrix}
  \delta &   0 & \frac{1}{2}\Omega(z)(1-e^{-\frac{t}{\tau_{r}}}) \\
  0 & 0 &  0 \\
 \frac{1}{2}\Omega(z)(1-e^{-\frac{t}{\tau_{r}}}) & 0 & -\delta
 \end{pmatrix}
\end{equation}
when the resonator is ringing up and the Hamiltonian 
\begin{equation}
H_{down}=\begin{pmatrix}
  \delta &   0 & \frac{1}{2}\Omega(z)e^{-\frac{t-t_0}{\tau_{r}}} \\
  0 & 0 &  0 \\
 \frac{1}{2}\Omega(z)e^{-\frac{t-t_0}{\tau_{r}}} & 0 & -\delta
 \end{pmatrix}
\end{equation}
when the resonator is ringing down. Quasi-static magnetic bath noise takes the form of a randomized detuning $\delta$ drawn from a Gaussian distribution with a standard deviation $\sigma=\sqrt{2}/T_2^*$~\cite{aiello2013}. The magnetic Ramsey measurement shown in Fig.~\ref{fig:sampleBRamsey} sets $T_2^*=0.68$~$\mu$s. 

Defining the result of this computation as the function $f(\tau_0,\Omega(z))$, we then perform a spatially-weighted average over the point spread function (PSF) of our confocal microscope to account for spatial inhomogeneities in our mechanical driving field. The resulting signal takes the form 
\begin{equation}
P_{\Ket{+1}}=\frac{C}{\int_0^\infty \! g(z,z_0) \, \mathrm{d}z}\int_0^\infty \! g(z,z_0)f(\tau_0, \Omega(z)) \, \mathrm{d}z
\label{eq:rabiHighQ}
\end{equation}
where $C$ accounts for partial polarization of the nuclear spin sublevel, $\Omega(z)=\Omega_{mech}|\text{sin}\frac{2\pi z}{\lambda_B}|$ is the mechanical driving field, $\lambda_B$ is the wavelength of the stress wave, and $g(z,z_0)$ describes a Gaussian approximation to a PSF centered at the focal depth $z_0$ with a depth dependent FWHM as described in Ref.~\cite{MacQuarrie2013}. To produce the model curve in Fig.~3b of the main text, we used the parameters $\Omega_{mech}/2\pi=3.8$~MHz, $z_0=5.9$~$\mu$m, $C=0.414$ (as measured via mechanically driven spin resonance), and $\lambda_B= 29.6$~$\mu$m. The simulation was repeated $200$ times, and these results were averaged to produce the final curve. 

\section{Ramsey Measurements}

\subsection{Ramsey Pulse Sequences}

\begin{figure}[htbp]
\centering
\mbox{\includegraphics[width=12cm]{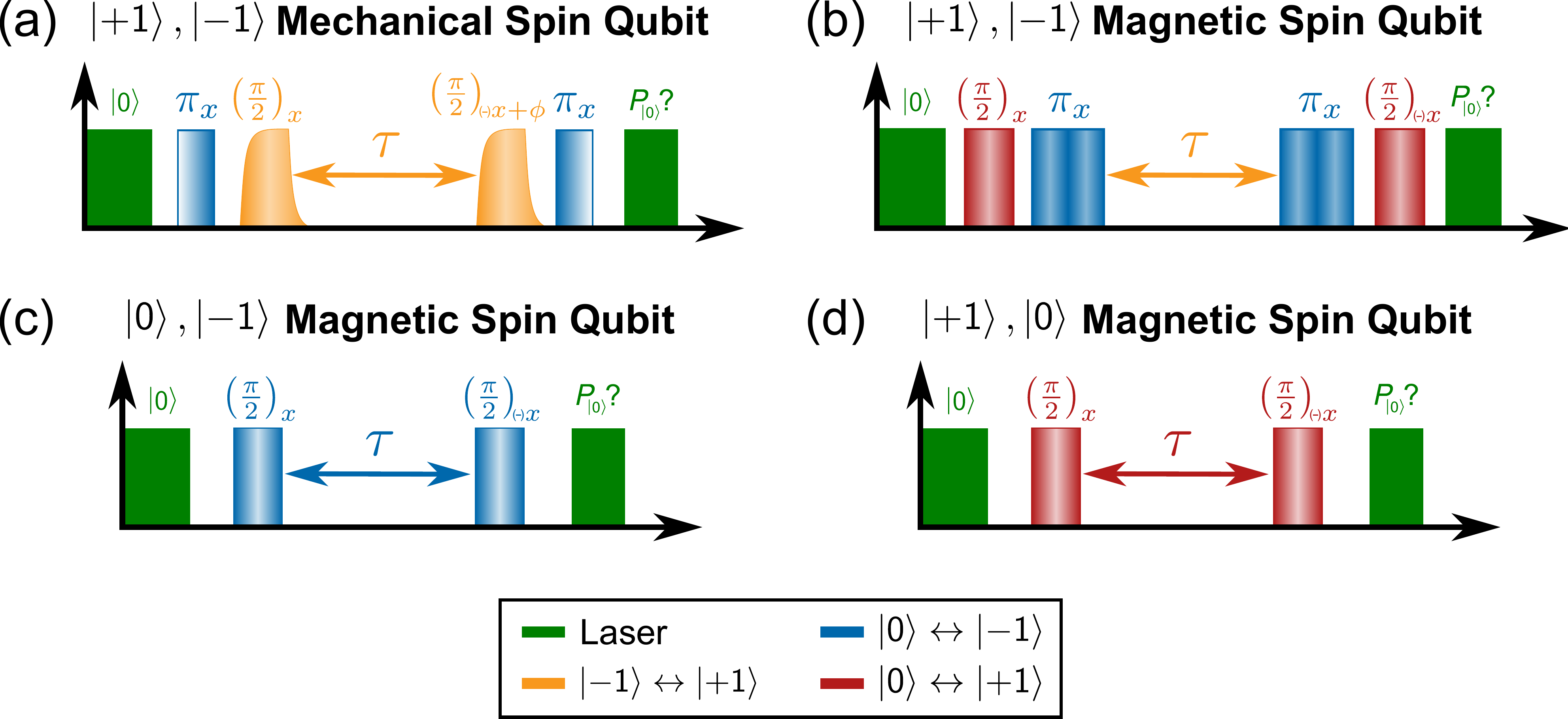}}
\caption{Pulse sequences used for the Ramsey measurements presented in the main text.}
\label{fig:ramseyPulse}
\end{figure}

Fig.~\ref{fig:ramseyPulse} shows the pulse sequences used for the Ramsey measurements presented in the main text. To eliminate experimental artifacts, we modified the typical Ramsey measurement to include a second measurement for each data point. We first execute the typical $\pi/2$---$\tau$---$\pi/2$ Ramsey sequence. Immediately afterward, we perform a $\pi/2$---$\tau$---$(-\pi/2)$ sequence. The difference of these two measurements equals twice the imaginary portion of the qubit's coherence $\text{Im}[\rho_{i,j}]$ ($i,j\in\{(m_s=)+1,0,-1\}$, $i\ne j$). We further modify the Ramsey sequence for the mechanically driven qubit by advancing the phase of the second $\pi/2$-pulse by $\omega_{rot}(\tau+\tau_{\pi/2})$. This extra phase shift introduces a known periodicity to the measurement that aids visualization of the decay envelope. 

\subsection{Ramsey Measurement Normalization}

Two measurements were used to normalize the spin contrast for the magnetic Ramsey measurements in the \plusZero and \zeroMinus subspaces. The maximum spin signal $y_{NP}$ is measured by optically pumping the NV center into $\Ket{0}$, shuttering the laser for the fixed dark time in which no pulses were applied, and then reading out the NV center fluorescence. Applying a single magnetic $\pi$-pulse to the relevant qubit during that dark time gives the minimum spin signal $y_{\pi}$. Defining the $\pi/2$---$\tau$---$\pi/2$ measurement results as $y_{+}$ and the $\pi/2$---$\tau$---$(-\pi/2)$ measurement results as $y_{-}$, the expression
\begin{equation}
\text{Im}[\rho_{ij}]=\frac{1}{2}\frac{y_{+}-y_{-}}{y_{NP}-y_{\pi}}
\end{equation}
gives the normalized coherence of the $\Ket{i},\Ket{j}$ qubit.

For the magnetic \plusMinus qubit Ramsey measurement, the same ``no pulse'' measurement gives the maximum spin signal $y_{NP}$. We define the minimum spin signal $y_{\pi}$ as the average of the signal from a single magnetic $\pi$-pulse on the \plusZero qubit and the signal from a single magnetic $\pi$-pulse on the \zeroMinus qubit. 

For the mechanically driven \plusMinus qubit, the ``no pulse'' measurement once again sets the maximum spin signal for the mechanically driven \plusMinus qubit. The minimum spin signal is set by a $\pi_{mag}$---$\pi_{mech}$---$\pi_{mag}$ pulse sequence. Here, $\pi_{mag}$ corresponds to a magnetic $\pi$-pulse on the \zeroMinus qubit, and $\pi_{mech}$ describes a mechanical $\pi$-pulse on the \plusMinus qubit. 

\begin{figure}[htbp]
\begin{center}
\begin{tabular}{c}
\includegraphics[width=12cm]{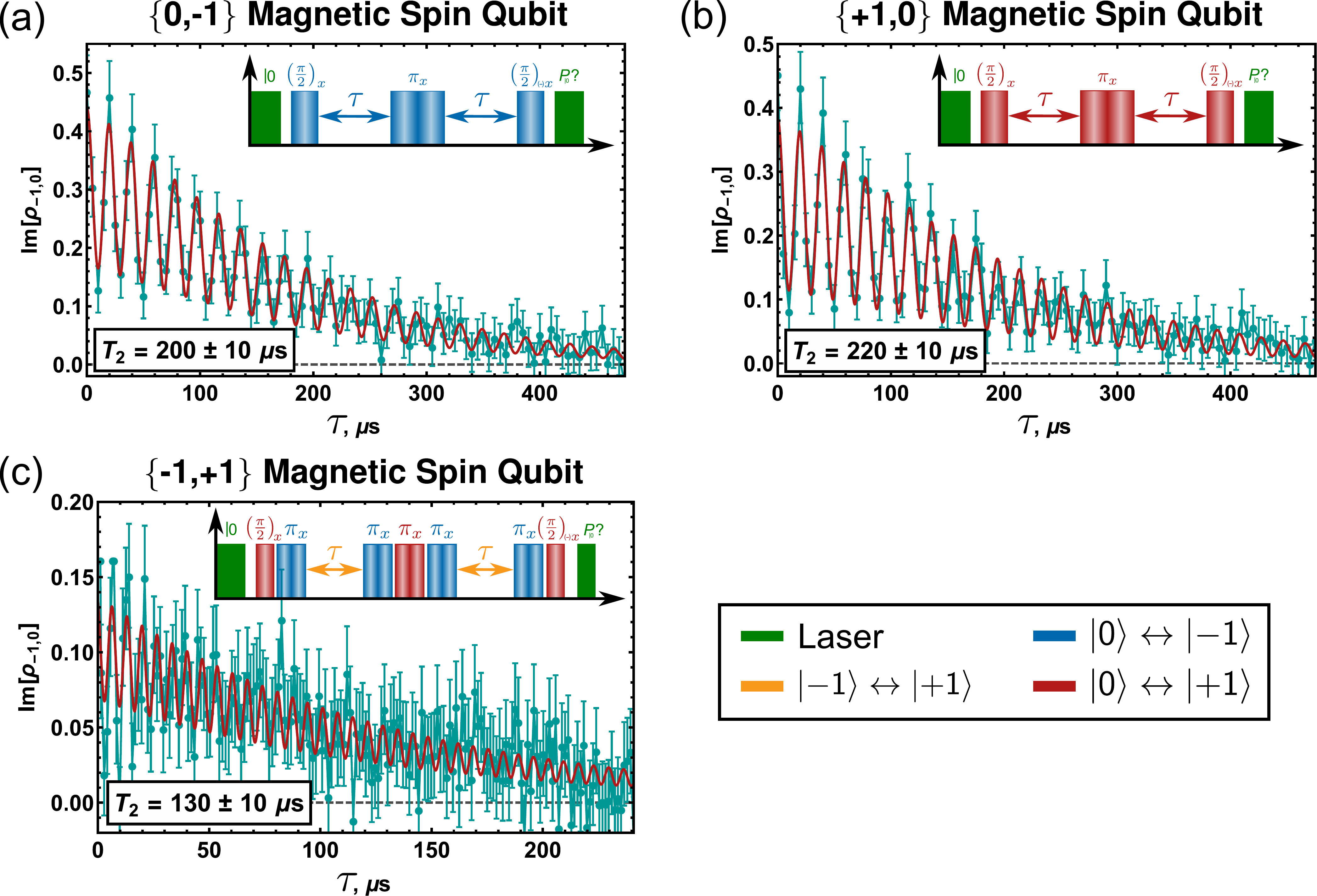} \\
\end{tabular} 
\end{center}
\caption[Hahn Echo Data] {Hahn echo data for (a) a magnetically driven \zeroMinus qubit, (b) a magnetically driven \plusZero qubit, and (c) a magnetically driven \plusMinus qubit. The pulse sequence for each measurement is inset within each plot.}
\label{fig:hahn}
\end{figure}

\section{Hahn Echo Measurements}

We performed magnetic Hahn echo measurements of the homogeneous dephasing time $T_2$ in Sample~A. We were unable to perform a mechanical Hahn echo experiment as intrinsic spin dephasing in our device limited the spin contrast after a mechanically driven $2\pi$ nutation to the prohibitive value of $\approx1$\%. Fig.~\ref{fig:hahn} shows the Hahn echo data for each magnetically driven qubit examined in the main text. Once again, we measure roughly twice the coherence for the \plusZero and \zeroMinus qubits when compared to the \plusMinus qubit.

\bibliography{bibDiMEMSarXiv}{}

\begin{thebibliography}{35}
\bibitem [{con()}]{controlRabiFootnote}%
  \BibitemOpen
  \href@noop {} {}\bibinfo {note} {In the SI, we also repeat this measurement
  reading out the population from the $\Ket{+1}$ state and observe similar
  behavior~\cite{SI}.}\BibitemShut {Stop}%
\end{thebibliography}


\end{document}